# A large-scale complexity-graded dataset of neuronal images and annotations

Wu Chen[1,4], Mingwei Liao[1,4], Xueyan Jia[2], Xiaowei Chen[2], Chi Xiao[3], Qingming Luo[3], Hui Gong[1,2], and Anan Li[1,2,3,*]

[1] *MOE Key Laboratory for Biomedical Photonics, Wuhan National Laboratory for Optoelectronics, Huazhong University of Science and Technology, Wuhan 430074, China*

[2] *HUST-Suzhou Institute for Brainsmatics, JITRI, Suzhou 215123, China*

[3] *State Key Laboratory of Digital Medical Engineering, Key Laboratory of Biomedical Engineering of Hainan Province, School of Biomedical Engineering, Hainan University, Sanya 572025, China*

[4]*These authors contributed equally.*

*\*Correspondence: aali@brainsmatics.org*

## Abstract

Accurate reconstruction of neuronal morphology is essential for classifying cell types and understanding brain connectivity. Recent advances in imaging and reconstruction techniques have greatly expanded the scale and quality of neuronal data. However, large-scale, standardized annotated datasets remain limited. Here, we present an open, multi-level neuronal dataset covering the whole mouse brain. Using a hierarchical strategy, we divided imaging data from 237 mouse brains into about 13,570,000 standardized blocks, classified into four levels of reconstruction difficulty. With the custom-developed reconstruction platform, we achieved high-precision three-dimensional reconstructions of 9,676 neurons at the whole-brain scale. This dataset will be made publicly available, providing a valuable resource for algorithm development and brain circuit modeling in neuroscience research.

## Introduction

Neurons are the basic functional units of the nervous system, and their complete morphological reconstruction is essential for understanding neural circuits [1-3]. Thanks to advances in labeling and imaging techniques [4-10], high-resolution three-dimensional whole-brain images of mice can now be obtained. These images allow the extraction of neuronal morphology that preserves spatial topology and provides a solid foundation for neural network modeling and morphological analysis [11-14].

In 2013, scientists achieved the first long-range tracing of neurons across the entire mammalian brain using fluorescence micro-optical sectioning tomography (fMOST) [15]. This groundbreaking work laid the foundation for whole-brain-scale neural circuit research. Winnubst et al. [11] employed an efficient platform to image and semi-automatically reconstruct the complete morphologies of over 1,000 projection neurons. Qiu et al. [16] focused on hippocampal neuron projections, reconstructing 10,100 neurons in the mouse hippocampus and creating an open online database. These studies not only reconstructed the three-dimensional morphologies of a vast number of neurons but also provided critical data for understanding the structure and function of the nervous system. However, in the face of rapidly growing massive imaging datasets, current neuron reconstruction still primarily relies on manual annotation, suffering from long reconstruction cycles and inconsistent quality. There is an urgent need to develop automated and standardized reconstruction methods.

In recent years, the application of artificial intelligence in neuron reconstruction has made significant progress and greatly improved the efficiency of image processing and recognition [17-20]. The performance of related reconstruction networks largely depended on large-scale, high-quality neuronal annotation data, which provided essential benchmarks for model training and validation [21]. In 2010, the DIADEM dataset was established, comprising six distinct datasets covering various animal species, brain regions, neuron types, and visualization methods [22]. While DIADEM's advanced algorithm development [23-30], its small scale and limited diversity constrained its utility [21, 31]. The BigNeuron project released a more extensive dataset, with its initial phase focusing on reconstructing sparsely labeled neurons. The project's "gold166" subset contained 166 three-dimensional neuronal images, each accompanied by corresponding manual reconstruction results, thereby significantly enriching the available annotated data resources [21]. However, this collection was limited to isolated neurons with relatively sparse distributions, lacking the complex fiber interference and densely packed neuronal arrangements found in biological systems [32]. As such, it failed to fully capture the remarkable complexity and diversity of neuronal morphology. While subsequent neuron datasets gradually accumulated substantial morphological reconstruction data [13, 33-39], these collections still proved inadequate for training high-precision reconstruction algorithms due to several key challenges. First, the reconstruction quality standards were inconsistent and failed to reach the gold-standard level of BigNeuron. Second, the lack of corresponding raw images prevented algorithms from establishing accurate mapping relationships between images and structures during training. Additionally, multiple factors affecting reconstruction quality, combined with significant variations in reconstruction difficulty across different brain regions, led to feature confusion when using such mixed datasets for training. As a result, models struggled to effectively capture the essential characteristics of neuronal morphology, ultimately causing unstable reconstruction performance. The HiNeuron dataset [33] successfully categorized data by difficulty and improved consistency, and it also helped enhance network performance. But remained limited in scale due to labor-intensive manual collection. Its download-only access model, lacking online query capabilities, hindered large-scale utilization.

In this study, our task was to build an open, annotated, and hierarchically organized mouse whole-brain dataset to meet the growing need for precise data in neuroscience. Using high-resolution optical microscopy, we performed three-dimensional imaging of the entire brain and captured detailed neuronal morphologies. With a combination of automated reconstruction and manual proofreading, we provided accurate neuron annotations. To process large-scale data, we developed an automated image tiling pipeline that divided raw images into about 13,570,000 sub-blocks. Using a complexity-graded strategy, we classified neuronal images into four levels to optimize data processing efficiency. Based on our reconstruction platform, we successfully reconstructed 9,676 neurons at the whole-brain scale. Through graded reconstruction and online collaboration, we ensured high-throughput and accurate annotation. The platform also offered API services for programmatic access and batch data downloads. By releasing this dataset, we aim to support whole-brain neuroscience research with more precise and comprehensive data. Our goal is to promote AI-based automated neuron reconstruction and enable high-quality, low-cost mapping of circuits from millions to tens of billions across the mammalian brain.

**Result**

**Neuronal image data acquisition pipeline**

The neuronal data acquisition pipeline consisted of four key phases: sample preparation, microscopic imaging, data preprocessing, and morphological reconstruction, as shown in Fig. 1. First, adeno-associated viruses were used to sparsely label neurons in specific mouse brain regions, reducing neuronal density in the imaging dataset. The brain samples were then embedded to create stable imaging specimens. Subsequently, the fMOST microscopy system acquired high-resolution three-dimensional images of entire mouse brains. Due to field-of-view limitations, the system captured data through sequential scanning, with individual scans being stitched together to reconstruct complete coronal sections. During imaging, uneven illumination introduced artifacts that complicated subsequent reconstruction, necessitating preprocessing to enhance image quality and minimize tracing errors.

Following preprocessing, neuronal tracing was performed on the optimized images. Given the massive scale of whole-brain datasets, the images were first divided into manageable blocks [40]. Automated tracing proceeded block-by-block until all neuronal fibers were processed. In the quality control phase, independent annotators performed initial tracings, which were then verified by domain experts to produce the final result of neuron reconstruction.

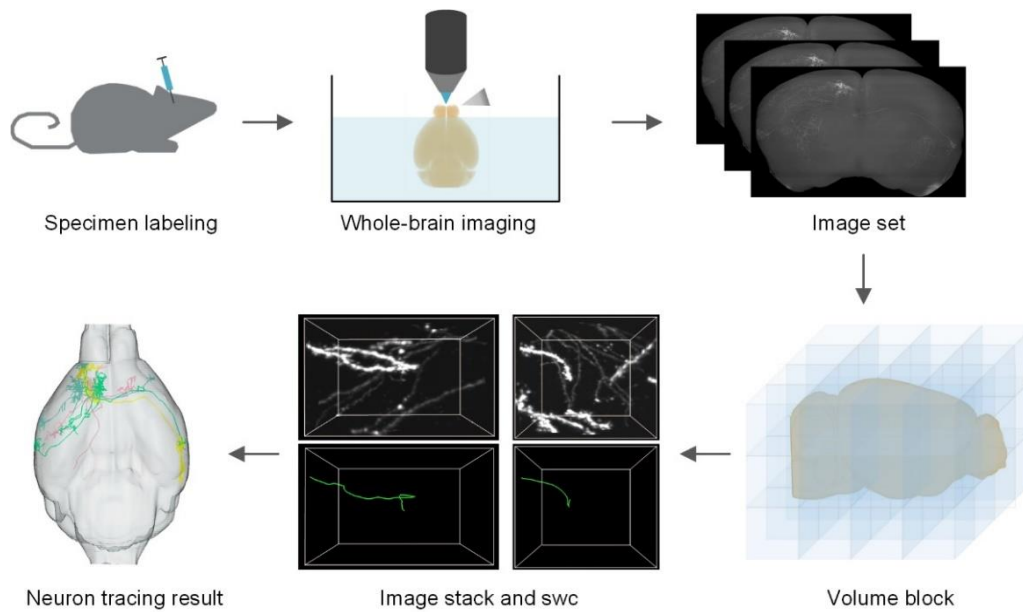

Figure. 1. Workflow of neuronal data acquisition. The process involved sample preparation, optical imaging, data preprocessing, and neuronal morphology reconstruction to obtain complete neuronal reconstruction data.

**A complexity-graded neuronal image dataset**

The neuron-based segmented images contained varying levels of complexity. In these images, the intricacy of fiber structures and their distribution density exhibited significant differences, making the difficulty of reconstruction tasks vary. If data of different complexity levels were analyzed together, it could become challenging to distinguish which data posed critical difficulties, thereby affecting the overall accuracy of the reconstruction results. By grading the refined data, it helped differentiate between data of varying difficulty levels.

To effectively classify the data, it was first necessary to define how to characterize the difficulty levels of neuronal images. This was achieved by employing various image processing techniques and computational metrics to quantify and describe the features of neural fibers within the images, thereby providing a basis for difficulty grading. During data classification, topological features, geometric features, spatial distribution features, and statistically derived features [21] were separately computed and integrated to distinguish different difficulty levels, as illustrated in Fig. 2a.

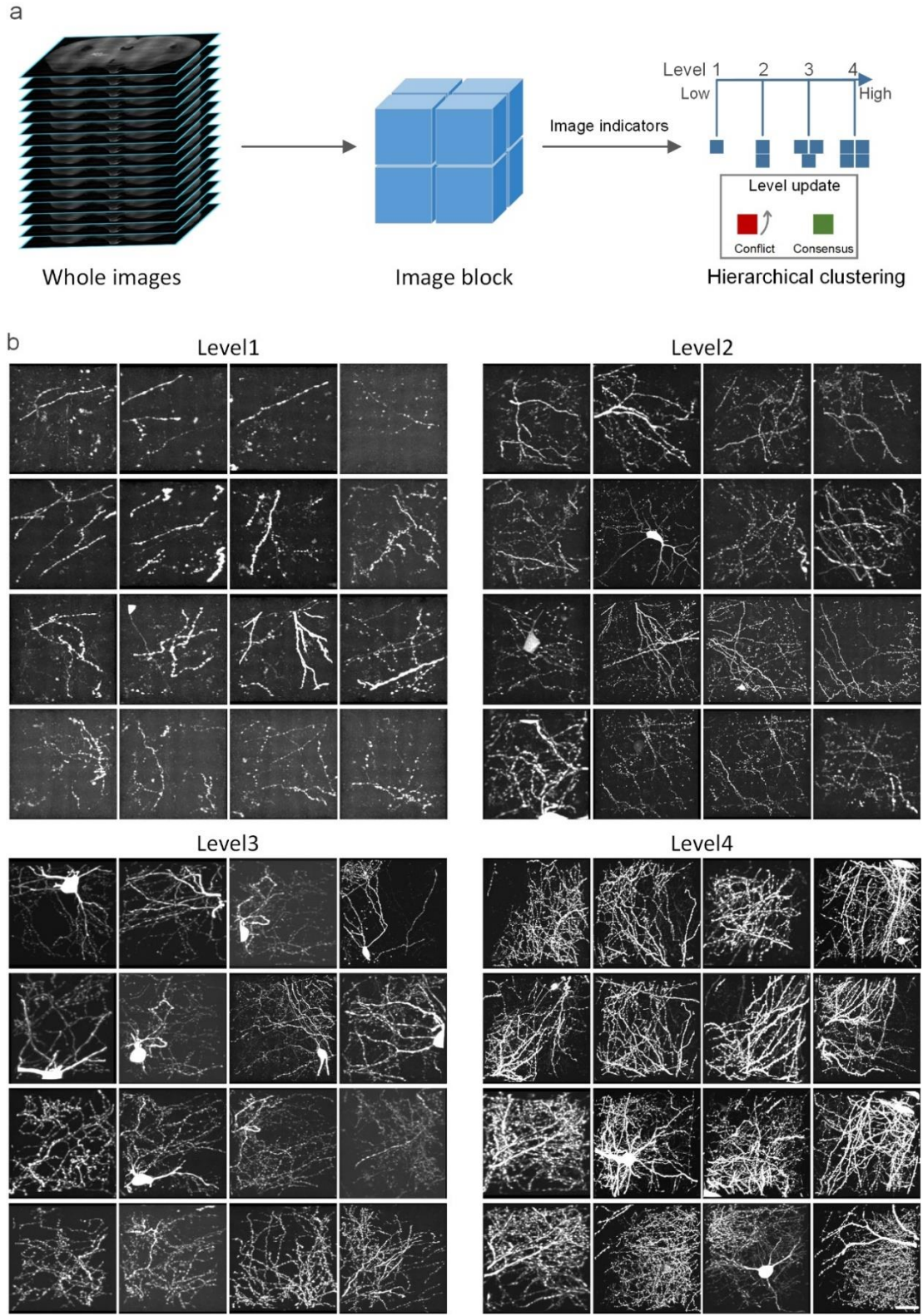

Figure 2. Hierarchical neuronal image data. (a) The workflow for neuronal data grading. (b) Results of neuronal data grading from Level1 to Level4. Scale bar, 50 μm.

First, a detailed analysis of the data containing complex fiber structures was performed. Due to the enormous size of whole-brain neuronal images, directly processing the entire image was computationally infeasible. Thus, an image tiling technique was adopted, where the original image was partitioned into smaller blocks according to predefined rules. These blocks served as independent data units, each representing a localized region spatially associated with neuronal fibers. The complexity of these regions could vary across different locations. Next, difficulty metrics for each image block were calculated to quantify regional characteristics, including fiber density, fiber length, keypoint density, and target fiber length.

Based on these computed metrics, a clustering algorithm was applied to classify the images, grouping blocks with similar difficulty into the same category. This resulted in neuronal images being stratified into distinct difficulty levels, effectively differentiating the reconstruction challenges across the dataset. Consequently, through data classification, the neuronal images were divided into four difficulty levels: Level1, Level2, Level3, and Level4, as shown in Fig. 2b. The data difficulty increased progressively from Level1 to Level4, where Level1 represented relatively easy images, Level2 denoted moderately difficult images, Level3 indicated more challenging images, and Level4 stood for the most difficult images. During actual neuron reconstruction, the data classification results were adjusted. If discrepancies arose in the reconstruction assessment for a particular block, its classification was updated. By grading the neuronal images, the complexity of each dataset was clarified, thereby identifying which data would pose greater challenges during the reconstruction process.

**Large-scale collection of neuronal morphology data and statistics**

This study used a custom-developed reconstruction platform to trace neuronal images. First, large-volume images were divided into smaller tiles. Each tile was automatically traced to generate initial fiber reconstructions, which were then loaded into the platform with their source images, as shown in Fig. 3a. During manual verification, operators adjusted brightness, rotated views, and zoomed in on the three-dimensional interface to evaluate tracing accuracy. Errors were corrected using editing tools like branch pruning, as shown in Fig. 3b. After processing each tile, the system automatically proceeded to the next until all tiles were completed.

After neuronal tile reconstruction was completed, the image tiles were stitched together to form a complete neuron reconstruction. Adjacent tile segments were spatially aligned, considering both three-dimensional coordinate consistency and morphological continuity of neural fibers to achieve seamless integration.

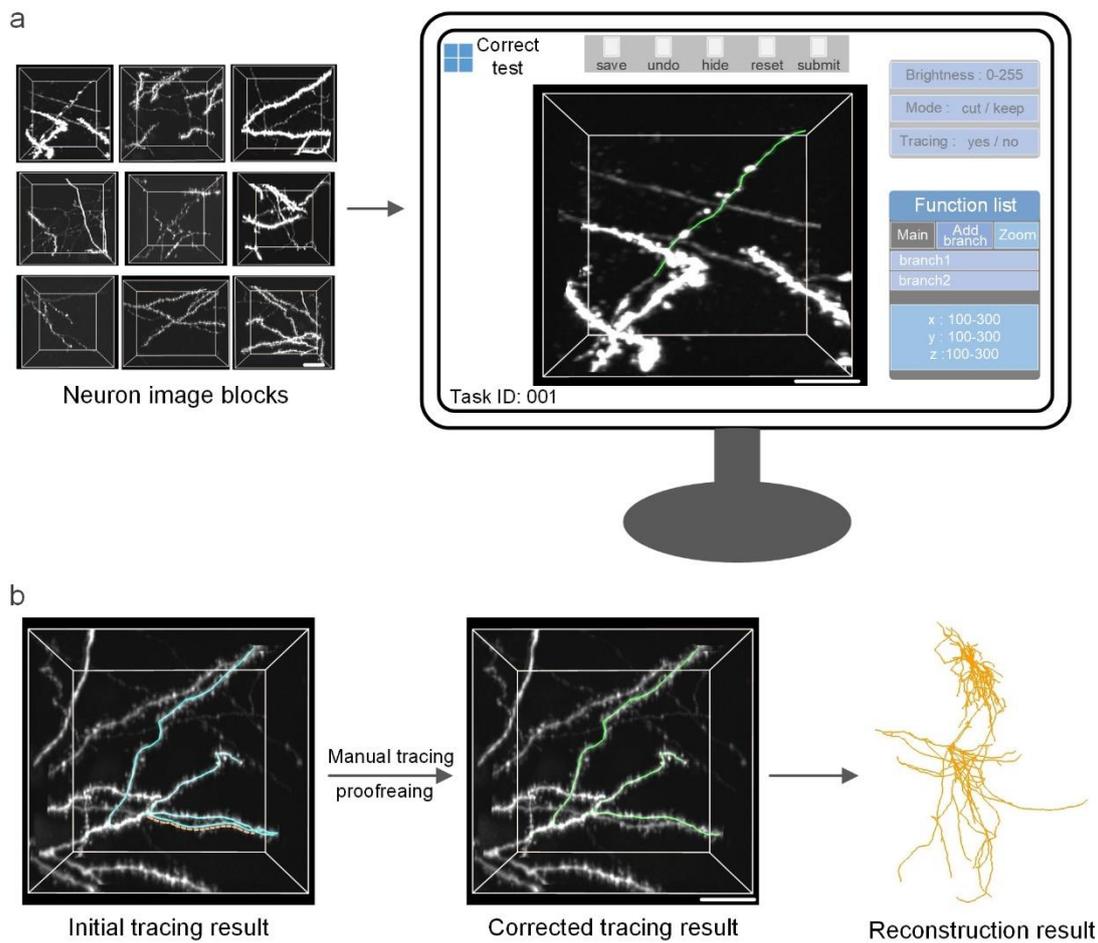

Figure 3. Workflow for Large-scale collection of neuronal morphology data. (a) The three-dimensional neuronal reconstruction platform. (b) Neuron reconstruction, correction, and results. Scale bars: a, b, 50 μm.

The system's data grading strategy automatically assigned tiles to operators based on complexity levels: experienced operators handled difficult, dense-branching regions requiring advanced spatial recognition and microscopy image interpretation skills, while junior operators processed simpler tiles with clear fiber paths and clean backgrounds. This human-machine collaborative approach significantly improved overall reconstruction efficiency.

The entire neuron reconstruction acquisition process balanced efficiency with precision. The automated reconstruction phase handled massive datasets, enabling rapid and relatively accurate reconstruction of simpler data while significantly improving processing throughput. Manual verification ensured reconstruction accuracy, particularly for precise reconstruction of error-prone algorithmic areas such as branches and crossings, thereby providing reliable foundational data for subsequent morphological analysis.

The study processed 237 mouse brain samples using a hierarchical strategy to classify massive microscopic image data into processing units with different complexity levels. The system handled a total of 13,569,022 neuronal image stacks, including 11,558,698 at Level1, 892,777 at Level2, 764,187 at Level3, and 353,360 at Level4, as shown in Fig. 4a. Through an integrated workflow combining automated processing and expert manual verification, the system successfully completed three-dimensional reconstruction of 9,676 neurons at a whole-brain scale, demonstrating its capability to process massive neural network datasets. Analysis of reconstructed neuron counts across samples revealed two categories: those containing 20 to 100 neurons and those exceeding 100 neurons, presented in Fig. 4b. More than 80% of samples yielded fewer than 60 reconstructed neurons, while only 6% of samples generated over 100 reconstructed neurons, indicating the system's adaptability to neural networks of varying scales.

To ensure reconstruction quality, the platform implemented a rigorous verification and correction protocol. The manual verification and correction counts for each data level revealed: Level1 required 5,116,512 corrections, Level2 needed 408,233, Level3 demanded 472,367, and Level4 accounted for 204,135 corrections. These statistics not only demonstrated the enormous scale of reconstruction work but also validated the effectiveness of the hierarchical strategy. More complex data blocks, though fewer in quantity, showed higher correction ratios: Level3 corrections represented 61.8% of its total data, while Level1 accounted for 44.3%. The successful processing of this massive dataset confirmed the platform's stability and reliability in large-scale neuronal reconstruction tasks.

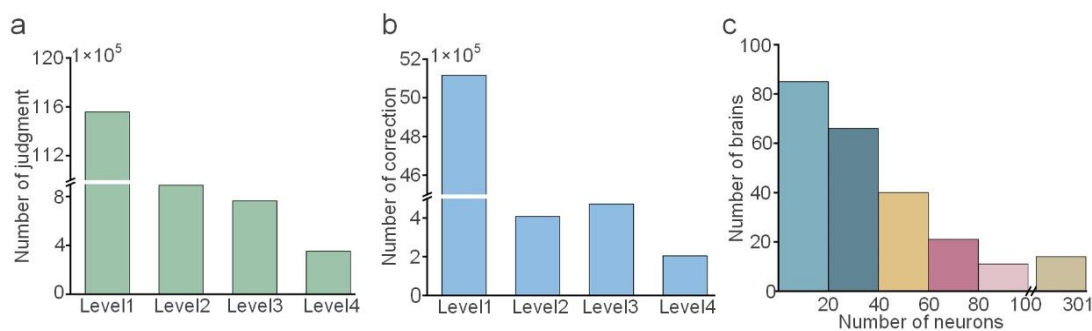

Figure 4. Neuronal image tile quantification and reconstruction statistics. (a) Image data statistics were judged in the reconstruction platform. (b) Corrected image data statistics in the reconstruction platform. (c) Whole-brain distribution of reconstructed neurons and per-sample neuron counts.

**The impact of neuronal image data on network training**

The neuronal reconstruction platform in this study handled massive whole-brain neuronal datasets, fully demonstrating its high-throughput processing capability. Furthermore, experiments were conducted to verify the impact of the amount of neuronal image data on network training performance and to analyze its role in improving neuron tracing accuracy. In the experiments, a 3D U-Net network [41] was employed, trained with 100, 400, and 1,000 three-dimensional neuronal image patches, respectively. The trained network was then used to extract neuronal signals, followed by neuronal tracing using the SparseTracer method [42]. To quantitatively evaluate the tracing performance, F1-scores were calculated across three data complexity levels (Level1, Level2, and Level3) to measure reconstruction accuracy.

As shown in Fig. 5, compared to direct tracing of raw images, signal extraction through the trained network significantly improved tracing accuracy. Furthermore, network performance demonstrated progressive improvement with increasing training data. Specifically, with 100 training samples, the network showed improved capability in neuronal structure extraction, yet improved the F1-score from 0.584 5 to 0.755 3. When the sample number increased to 400, the F1 score rose to 0.782 4. With 1 000 training samples, the F1 score further increased to 0.803 9, enabling more accurate reconstruction of complex neuronal structures, indicating that larger data facilitated learning of more robust feature representations. Performance improvements were also observed for Level2 and Level3 data. These experiments confirmed the positive correlation between data scale and model performance, while revealing performance variation across different complexity levels. The experimental results provided important references for optimizing neuronal reconstruction algorithms.

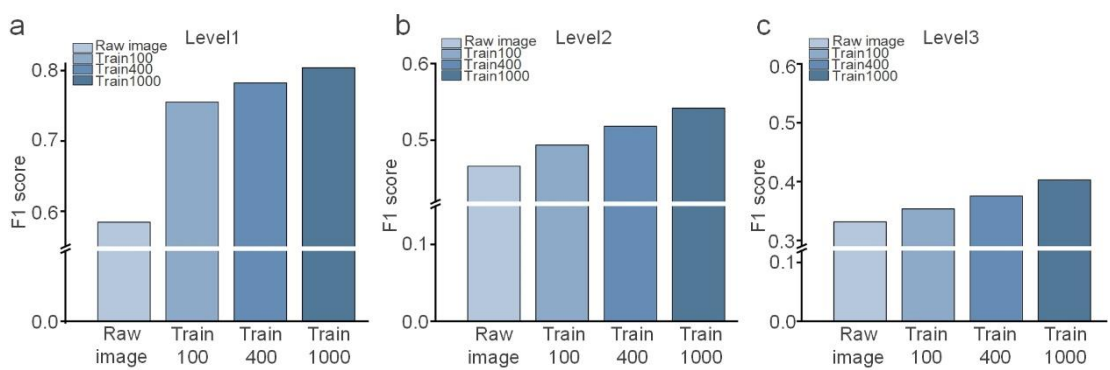

Figure 5. Quantitative results of model performance with different numbers of training data. (a) Performance of models trained with different data in Level1. (b) Performance of models trained with different data in Level2. (c) Performance of models trained with different data in Level3.

**Analysis of reconstruction difficulty for neuronal axon and dendrite data**

During the complete neuronal reconstruction process, significant structural differences were observed between axons and dendrites. Axons exhibited longer branching patterns that frequently spanned multiple brain regions, while dendrites generally displayed shorter branches primarily localized near the cell body [43]. To systematically evaluate whether these structural differences translated into varying reconstruction difficulties, we conducted a comprehensive statistical analysis comparing the difficulty-level distributions between axonal and dendritic reconstructions. This analysis enabled the development of structure-specific reconstruction protocols. The upper three panels of Fig. 6a present detailed axonal reconstruction statistics. The first panel shows the initial difficulty distribution of axonal data, with Level1 comprising 66.3% of cases, Level2 accounting for 26.5%, Level3 representing 6.2%, and Level4 making up 1%. The second panel displays the distribution of corrected axonal data across difficulty levels, with proportions of 60.6%, 30.0%, 8.1%, and 1.3% for Level1 through Level4 respectively. The third panel reveals the correction conversion rates for axonal data, demonstrating a progressive increase from 59.4% at Level1 to 98.7% at Level4. Corresponding dendritic statistics are presented in the lower three panels of Fig. 6a. Comparative analysis of these distributions showed remarkably similar patterns between axonal and dendritic reconstructions across all measured parameters: initial difficulty proportions, correction ratios, and conversion rates. This consistency in difficulty-level distributions was particularly noteworthy given the fundamental structural differences between these neuronal components. The findings suggest that while axons and dendrites differ substantially in their morphological characteristics, they present comparable patterns of reconstruction challenges when analyzed through our hierarchical difficulty framework. These results provide valuable insights for optimizing reconstruction algorithms and resource allocation in large-scale neuronal mapping projects.

  The statistical results revealed distinct distributions of neuronal data across different difficulty levels, with lower-level data constituting a significantly higher proportion. This distribution pattern enabled efficient allocation of simpler reconstruction tasks to less experienced personnel, effectively lowering the technical threshold and expanding the potential workforce pool. Such strategic task assignment substantially accelerated large-scale neuronal reconstruction efforts. For the more challenging reconstruction components, the hierarchical classification system successfully identified and isolated these difficult cases from the larger pool of lower-level data. These complex reconstructions were then assigned to highly experienced operators, ensuring the accuracy and reliability of the final neuron reconstructions.

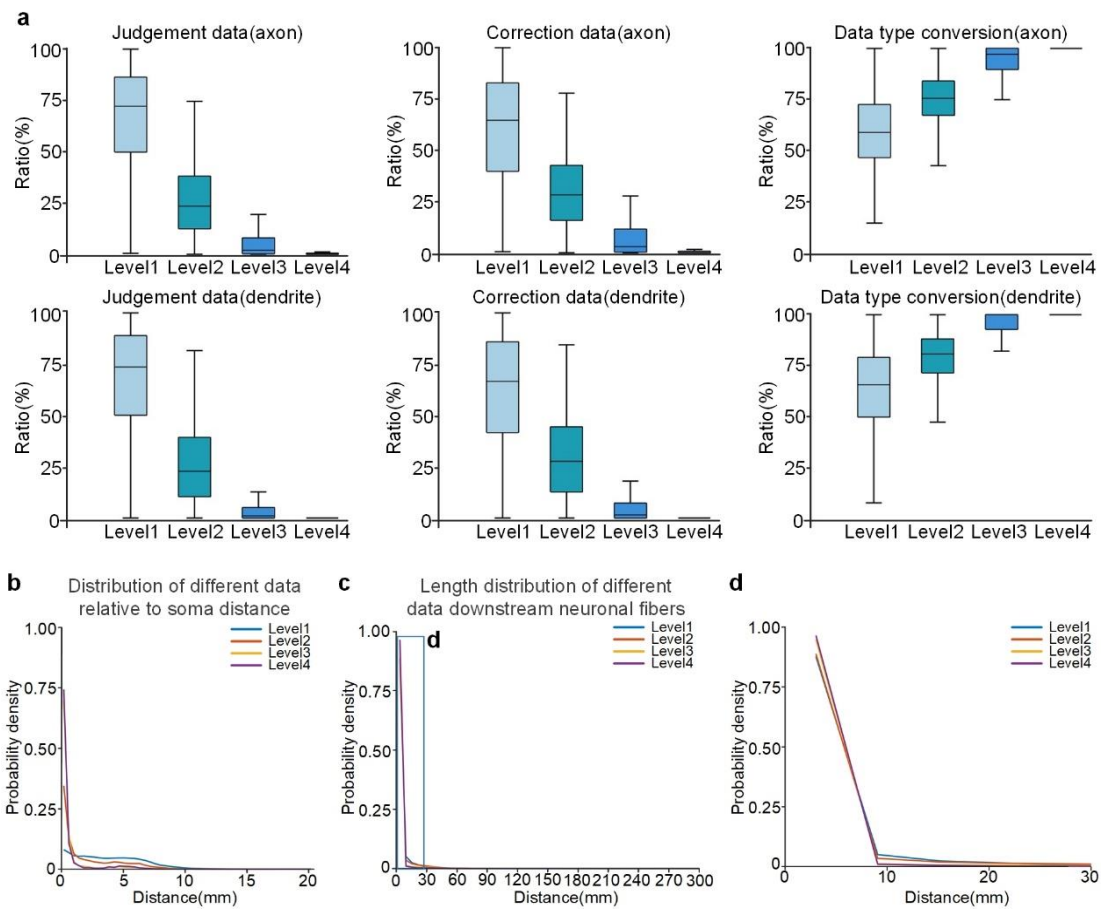

Figure 6. Difficulty differences in axonal and dendritic reconstruction. (a) Quantitative statistics of data at different difficulty levels for axons and dendrites. (b) Probability distribution of data quantities across difficulty levels relative to soma distance. (c) Proportions of different difficulty levels versus neuronal reconstruction length. (d) Variation in higher-difficulty data proportion with increasing neuronal length.

The study examined how neuron reconstruction difficulty varied by distance from the soma. As shown in Fig. 6b, higher-difficulty Level3 and Level4 data clustered closer to cell bodies, revealing greater complexity in these perisomatic regions. Some challenging cases still appeared in distant areas, demonstrating that difficulty existed throughout the neural tissue. In contrast, simpler Level1 reconstructions were evenly distributed within 10 mm of somata. Level2 data appeared most frequently at intermediate distances while still maintaining presence across the measured range.

To investigate whether the proportion of different difficulty levels correlated with neuronal length, we analyzed the relationship between reconstructed neuron length and difficulty-level distribution, as shown in Fig. 6c. The results demonstrated variations in difficulty proportions for neurons shorter than 30 mm. The magnified view in Fig. 6d revealed that the proportion of higher-difficulty data gradually decreased as neuronal length increased, indicating an inverse relationship between neuron length and reconstruction difficulty.

**Discussion**

Neurons serve as the fundamental functional units of the brain. Acquiring neuronal morphology is crucial for neuron classification, elucidating the operational mechanisms of neural circuits, and gaining deeper insights into information transmission and processing between brain regions. Leveraging viral neuronal labeling techniques and fMOST imaging technology combined with neuronal reconstruction workflows, we obtained whole-brain-scale neuronal images and their reconstruction results. This study established a hierarchical strategy for neuronal images, successfully constructing a large-scale dataset comprising thousands of neurons. This achievement lays an important foundation for the systematic investigation of neuronal morphological diversity and circuit functionality.

The study's core contribution was the establishment of a standardized, scalable neuronal reconstruction data framework. Through a hierarchical processing strategy, we systematically categorized massive neuronal image data (13 569 022 images ) into four complexity levels and optimized reconstruction workflows for each tier, significantly improving both processing efficiency and accuracy. Critically, the resulting dataset encompassed not only raw image stacks and reconstruction outputs but also complete morphological data for individual neurons. This structured data organization enabled dual functionality, serving as training resources for machine learning algorithms while simultaneously supporting multiscale morphometric statistical analyses. The framework thus delivered dual value for both methodological development and applied neuroscience research.

Future applications of this dataset hold significant potential in three key directions. First, continuous expansion of sample sizes will enable the development of more comprehensive neuronal morphology benchmark sets, providing increasingly reliable gold-standard data for reconstruction algorithm evaluation. Second, the dataset's structured architecture makes it particularly suitable for training next-generation intelligent reconstruction algorithms, especially for developing specialized processing modules targeting regions of varying complexity. Finally, its standardized processing pipeline can serve as a reference for neuron reconstruction studies in other model organisms. While the current dataset already encompasses substantial neuronal diversity, further increasing both sample sizes and brain region coverage will enhance its utility as a computational neuroscience tool. Such expansions will advance neuronal digital reconstruction technologies toward the dual goals of higher throughput and greater precision, ultimately accelerating progress in large-scale neural circuit mapping.

**Methods**

**Whole-brain optical imaging**

All histology processing steps were followed by established protocols [8, 10]. We acquired whole-brain images of embedded mouse brains using the fMOST system. The embedded samples underwent high-precision mechanical sectioning while simultaneously capturing surface images. After completing optical imaging of each layer, the surface tissue was removed, and this cycle was repeated until the entire brain was covered. This approach overcame optical depth limitations, enabling continuous three-dimensional tomography of centimeter-scale specimens (e.g., mouse whole brains). The raw images were then stitched into complete coronal sections, with illumination correction applied to enhance image quality for subsequent processing and analysis. Our dataset achieved voxel resolutions of either 0.35 μm × 0.35 μm × 1 μm or 0.32 μm × 0.32 μm × 1 μm, stored in 16-bit depth LZW-compressed TIFF format.

**Neuronal three-dimensional reconstruction platform**

We developed a neuronal reconstruction platform that enabled high-throughput, high-quality whole-brain neuron reconstruction through a hierarchical collaborative approach. The SparseTracer automated algorithm first reconstructed neuronal images [42], followed by back-to-back manual verification to generate gold-standard annotations. When reconstruction errors were identified, manual corrections were performed. Each reconstruction underwent validation by at least two independent annotators to ensure back-to-back verification consistency. In cases of inter-annotator disagreement, a senior expert made the final determination to resolve conflicts.

We implemented a collaborative workflow on an interactive cloud platform, where multiple users jointly verified neuron image tiles pre-reconstructed by automated algorithms. This approach significantly reduced reconstruction difficulty while substantially improving neuron reconstruction throughput. Through our data-tiering strategy, we precisely allocated reconstruction tasks of varying difficulty levels to appropriately skilled users. A scoring model dynamically tracked reconstruction accuracy, ensuring consistent output quality across all tiers.

**Hierarchical organization of neuronal images**

The neuronal reconstruction dataset constructed in this study systematically integrated whole-process data ranging from raw image processing units to complete neuronal morphologies. For neuron image patches, we established a comprehensive "mouse brain-image patch" indexing system. These image patch data not only contained original microscopic optical images (each patch measuring several hundred pixels in size) but also provided manually verified reconstruction results saved in SWC format [44]. Based on the difficulty levels of neuronal morphologies, we categorized all neuron image patches into four processing grades: Level1 contained relatively simple fiber structures, Level2 covered areas with moderate fiber distributions, Level3 comprised more challenging complex regions, and Level4 represented the most difficult fiber structures.

This data organization approach preserved the detailed features of the original image processing units while providing integrated complete neuronal morphology information, establishing a multi-scale research platform for neuroscience studies, from microscopic reconstruction details to macroscopic morphological features. All data were uniformly standardized in their descriptions and could support efficient retrieval and three-dimensional visualization.

**Metadata**

The dataset in this study documented detailed information about the acquired mouse samples. Each sample included the mouse brain ID, along with the strain, sex, and age of the mouse. The imaging parameters covered the microscope system used, the number of imaging channels, and the imaging resolution. Specifically, the dataset contained the following details:

Brain ID: The unique identifier for each mouse brain, used for data management and retrieval.

Line: The genetic background of the mouse sample (e.g., C57BL/6J, PV-Cre).

Gender: The sex of the mouse (male or female).

Age: The age of the mouse was recorded in weeks.

Labeling strategy: The labeling method used for the sample.

Imaging system: The imaging equipment employed (e.g., HD-fMOST, TDI-fMOST).

Imaging channel: The optical channels used during imaging (e.g., GFP, YFP).

Voxel resolution: The resolution of the acquired three-dimensional image data.

Taking the imaging systems as an example, we detailed the neuronal image dataset information. The HD-fMOST imaging system included 191 mouse brain samples, with the acquired neuronal images having a three-dimensional resolution of either 0.32 μm × 0.32 μm × 1 μm or 0.32 μm × 0.32 μm × 2 μm. The TDI-fMOST imaging system contained 46 mouse brain samples, all with a uniform three-dimensional resolution of 0.35 μm × 0.35 μm × 1 μm. Each neuronal image had a three-dimensional size of several hundred pixels, and every image patch was accompanied by a corresponding reconstructed SWC file.

This dataset provided comprehensive data support for neuronal morphology reconstruction, automated image processing, and related algorithm development, while offering abundant foundational data resources for subsequent large-scale neural morphology analysis and intelligent processing.

**Technical validation**

During neuronal image preprocessing, we implemented a rigorous preprocessing pipeline to ensure the integrity of the acquired mouse brain neuronal image data. First, the imaging system captured raw strip-shaped images of mouse brain sections with intentional spatial overlap during acquisition. This overlapping configuration effectively prevented structural discontinuities or misalignments at stitching junctions. Subsequently, we employed a feature-based automatic image stitching algorithm to precisely align these overlapping regions, generating seamless and continuous whole-brain images. This stitching process adhered to spatial continuity principles, ensuring structural coherence of the stitched brain images in both two-dimensional and three-dimensional spaces. Through systematic quality checks, all stitched regions achieved high spatial consistency and accuracy. These preprocessed images established a high-quality foundation for subsequent neuronal structure analysis.

During neuron reconstruction, we implemented a multi-operator cross-verification method to ensure reconstruction accuracy. First, two independent reconstruction operators performed separate validity assessments on the same data block. When their judgments matched, we calculated the reconstruction accuracy rate for that block. If the rate reached the predefined threshold, the reconstruction result was validated as correct and adopted as the reference standard. If the initial two assessments disagreed, a third operator was introduced for further evaluation, and a comprehensive accuracy rate was recalculated based on all three judgments. When the accuracy rate met the predefined standard, the reconstruction result was confirmed. If the threshold remained unsatisfied, a fourth operator was engaged, with this iterative process continuing until the target accuracy rate was achieved. This tiered cross-verification strategy ensured the reliability and precision of neuronal image reconstruction results.

**Usage notes**

The entire neuronal image dataset contained multiple difficulty levels, each comprising distinct neuron image blocks. To facilitate optimal dataset utilization, we documented the access, visualization, and usage methodologies.

First, users could access the dataset through our official website. The platform enabled browsing of detailed information for each mouse brain data sample, including strain specifications, imaging systems, and resolution parameters. Subsequently, users could view the data by selecting the View option.

Upon entering the image visualization interface, the left-side menu bar provided a data selection function categorized by difficulty levels, allowing users to browse image blocks of different grades as needed. In the main view area of the page, the selected neuron images and their corresponding fiber reconstruction results were displayed. Additionally, users could adjust image brightness using built-in page controls and zoom in/out flexibly with the mouse wheel to examine detailed structures.

To support local data processing, users could perform efficient data downloads through our provided download options. Each sample featured a "Download" button, enabling users to download data on a per-sample basis. The downloaded compressed files were organized in a multi-level directory structure: "Brain ID/Neuron/image|swc|txt". The first-level directory "Brain ID" represented the sample identifier, while the second-level directory "Neuron" contained the reconstructed neurons for that sample. The third level consisted of three subdirectories: "image" stored the three-dimensional neuron image blocks, "swc" held the corresponding reconstructed neuron results for each image block, and "txt" maintained the classification information for each image block. This structured data organization allowed users to quickly locate and process required information, significantly enhancing local analysis efficiency.

## Acknowledgements

We would like to thank the MOST group of Britton Chance Center for Biomedical Photonics, Wuhan National Laboratory for Optoelectronics, MOE Key Laboratory for Biomedical Photonics, Huazhong University of Science and Technology. This study was supported by STI 2030-Major Projects (2021ZD0201002) and the National Natural Science Foundation of China Grants (T2122015, 32192412).

## Author contributions

Hui Gong, Qingming Luo and Anan Li conceived and designed this study. Wu Chen, Mingwei Liao, and Chi Xiao performed the experiments, data analysis, and graph drawing. Wu Chen, Mingwei Liao, Xueyan Jia and Xiaowei Chen prepared the neuronal data. Wu Chen, Mingwei Liao, Chi Xiao and Anan Li wrote this paper.

## Competing interests

The authors declare no competing interests.